\documentclass[namedreferences]{SolarPhysics}
\usepackage[optionalrh]{spr-sola-addons} 
\usepackage{graphicx}        
\usepackage{color}           
\usepackage{url}             

\newcommand{\etal}{{\it et al.}}

\newcommand{\aap}      {{\it Astron. Astrophys.}}

\newcommand{\apj}      {{\it Astrophys. J.}}

\newcommand{\solphys}  {{\it Solar Phys.}}

\begin{document}

\begin{article}

\begin{opening}
\title{Azimuth ambiguity removal and non-linear force-free extrapolation of near-limb magnetic regions \\ {\it Solar Physics}}

\author{G. V.~\surname{Rudenko}$^{1}$\sep
        I. I.~\surname{Myshyakov}$^{1}$\sep
        S. A.~\surname{Anfinogentov}$^{1}$\sep
       } \runningauthor{G.V.
Rudenko and I.I. Myshyakov} \runningtitle{Ambiguity and Force-free
field reconstruction near-limb}

   \institute{$^{1}$ Institute of Solar-Terrestrial Physics SB RAS, Lermontov St. 126,
Irkutsk 664033, Russia
                     email: \url{rud@iszf.irk.ru} email: \url{ivan_m@iszf.irk.ru} email: \url{anfinogentov@iszf.irk.ru}\\}
\begin{abstract}
Possibilities in principle for satisfactory removal of the
$180^{\circ }$-azimuthal ambiguity in the transverse field of vector
magnetograms and the extrapolation of magnetic fields independently
of their position on the solar disk are shown. Revealed here is an
exact correspondence between the estimated field and the
nonpotential loop structure on the limb.

The Metropolis's algorithm modified to work in spherical geometry is
used to resolve the azimuthal ambiguity. Based on a version of the
optimization method from \inlinecite {cite01}, we use corrected
magnetograms as boundary conditions for magnetic field extrapolation
in the nonlinear force-free approximation.

\end{abstract}
\end{opening}

\section{Introduction}
     \label{Section1}
Removing azimuth ambiguities is a necessary condition for using
vector magnetograms to detect a real magnetic structures and
understand the nature of active solar regions. At present we have
many methods and approaches providing a successful solution to this
problem but only for regions nearby the disk center or at a medium
distance from it (\inlinecite {cite02}; \inlinecite {cite03};
\inlinecite {cite04} \inlinecite {cite05}; \inlinecite {cite06};
\inlinecite {cite07}; \inlinecite {cite08}; \inlinecite {cite09};
\inlinecite {cite10};). For this reason near-limb regions are not
available for effective use. Thus a considerable part of collected
and constantly reproduced valuable solar data, most searched in
various investigations related to region physics and prediction of
their activity, is lost. The key problem of many methods for
resolving azimuth ambiguities (including the minimum energy method
applied in this work) is to construct correctly a near-boundary
model of potential or linear force-free fields answering strictly to
the longitudinal component of a photospheric field being measured.
The solution to this problem is quite simple and successful only for
central regions. Both getting a near-boundary model of potential or
linear force-free field and the rest of the technique for removing
azimuth ambiguities are to level most of the negative influence
associated with necessary flat approximation of the spherical part
of a region. For a number of reasons, this negative influence is
most pronounced in distal regions. In this regard, any examples of
successful removing of a near-limb vector magnetogram azimuth
ambiguity and subsequent reproduction of a real near-limb 3D
structure from a prepared complete boundary vector field are of
fundamental importance.

In this paper, we propose a method for solving the azimuth ambiguity
problem for near-limb regions and test it on model and real data. In
case of model data, correctness of the azimuth ambiguity removal can
be easily verified. Otherwise, in case of real data it can be
estimated only indirectly.  We make such estimation by performing
NLFF extrapolation, relying on corrected vector magnetogram, and
comparing extrapolated field lines to real loop structures. For NLFF
extrapolation we use the optimization method \inlinecite {cite07} as
implemented in \inlinecite {cite01} (FOM). Using this version, the
authors of \inlinecite {cite01} managed to reproduce a non-potential
sigma field structure observed at a medium distance from the center.
Clearly, the correctness of real field extrapolation depends first
of all on the correctness of boundary transverse field
disambiguation. On the other hand, the evidence of successful
solution to the problem of ambiguity may be a correspondence between
the extrapolated field and observed coronal features. Actually, we
use the studied example to conduct an interrelated test of two
techniques aimed at solving various problems: azimuth ambiguity
removal and NLFF extrapolation as applied to near-limb regions.

FOM of NLFF extrapolation is described in detail in \inlinecite
{cite01}. Here it is applied in its original form, without
modifications. The method for removing azimuth ambiguity rests on
the minimum energy method \cite {cite02}. This paper presents the
following components of this method:

- translation of data in the form of artificial Stokes parameters
into the working "quasi-spherical" coordinate system \cite {cite01}
succeeded by  smoothing to reduce noise component of the transverse
field and with the inverse transformation to the vector form;

- FFT extrapolation of the boundary potential field with constant
direction of the oblique derivative corresponding to the observed
line-of-sight component in the "quasi-spherical" coordinate
system;

- modification of the minimum energy method to spherical geometry
with no need for data grid uniformity ("spherical" minimum energy
method).

\section{Preliminary transformations of initial data}
      \label{Section2}
\subsection{Transformation of initial vector magnetogram data into artificial Stokes parameters}
  \label{Section2.1}
A data set specified on the uniform grid of rectangle of the picture
plane will be taken as an initial magnetogram: $B_{x}$ is the
line-of-sight magnetic field component; $B_{y}$ and $B_{z}$ are
transverse field components with the $180^{\circ }$ ambiguous
azimuth $\varphi $. Let us define the transformation of initial data
into new variables of artificial Stokes parameters:

\begin{eqnarray}\label{eq1}
  V = B_{x}, \qquad  Q = B_{\perp }\cos \left( 2\varphi \right), \qquad
U = B_{\perp }\sin \left( 2\varphi \right), \nonumber \\
\qquad \\
 B_{\perp }=\sqrt{B_{y}^{2}+B_{z}^{2}},\qquad \varphi
=\arctan \left( B_{z}/B_{y}\right). \nonumber
\end{eqnarray}

All $V, Q, U$ values are unabiguious. Using Equation (\ref{eq1}), we
can always return to field variables which not necessarily coincide
with initial transverse components. We can apply interpolation and
smoothing operations to artificial Stokes parameters (in contrast to
the indefinitely directed transverse field) without loss of physical
connections between transverse field components. For instance, we
can interpolate data without quality loss into another grid,
including nonuniform one, smooth them and then return to the field
form.

\subsection{Transformation of data into a coordinate system of the quasi-spherical space}
  \label{Section2.2}
Azimuth disambiguation method and the NLFF extrapolation require
transforming data into a uniform grid of the lower plane of the
square box in the quasi-spherical space. The coordinate system of
the quasi-spherical space working area relies on the auxiliary
coordinate system where a conditionally chosen center of magnetogram
is at the solar disk center $[R_{Sun},0,0]$. Longitude and latitude
in radians define horizontal coordinates in the quasi-spherical
space. This coordinate system is thoroughly described in \inlinecite
{cite01}. Next we consider the introduced coordinate system as
Cartesian in which the model field divergence is zero and the strict
condition of force-free approximation is valid. A desired boundary
field is specified as follows:

 - a uniform grid is plotted on a quasi-spherical surface with a spatial resolution scale less
 than the minimum scale of the data grid spatial resolution;

- initial magnetogram parameters V, Q and U of Equation (\ref{eq1})
are interpolated into nodes of the uniform grid of the working space
lower plane and, if necessary, the obtained data are smoothed and
interpolated into the desired uniform grid;

- data are transformed into the normal vector form;

- vector transformations of the field into the quasi-spherical
coordinate system are possible only after the azimuth ambiguity
problem is solved.

\section{Potential field calculations}
      \label{Section3}

After having transfered data to the working boundary plane, we can
use only a line-of-sight field component ($B_{x}$ - in the initial
coordinate system). To compute the potential field, we obtain
representation of the line-of-sight unit vector corresponding to the
magnetogram center in the working coordinate system. Next, in FFT
format we solve the trivial boundary Laplace problem with the
oblique derivative of constant direction. We use the potential field
in the spherical minimum energy method and to construct an initial
field in the 3D box for FOM.

Note that a similar geometric scheme can be easily applied to solve
the boundary problem of the force-free extrapolation with a given
constant $\alpha $

\section{Modification of the minimum energy method (spherical minimum energy method)}
      \label{Section4}
The minimum energy method \inlinecite {cite02} implements the
problem of searching a state corresponding to the global minimum of
the positively defined additive function of $\nabla \cdot {\bf B}$
and $J_{n}$ simulating energy characteristic of the statistical
system. The problem-solving method rests on the well-known and
widely used method simulating the annealing process (\inlinecite
{cite11}; \inlinecite {cite12}; \inlinecite {cite13})and, as applied
to the azimuth ambiguity problem, is discussed in detail in
\inlinecite {cite02}. Our modification of this method \inlinecite
{cite02} concerns only changes in the technique of determining local
values  $\nabla \cdot {\bf B}$ and $J_{n}$ and configuration of an
elementary micro-ensemble specifying "temperature" characteristic.
This modification simplifies the technique of determining  $\nabla
\cdot {\bf B}$ and $J_{n}$ on the sphere and can work equally well
on arbitrary nonuniform grids, including the one that corresponds to
uniform distribution of node projections onto picture plane.

\begin{figure}
\centerline{\includegraphics[width=1.\textwidth,clip=]{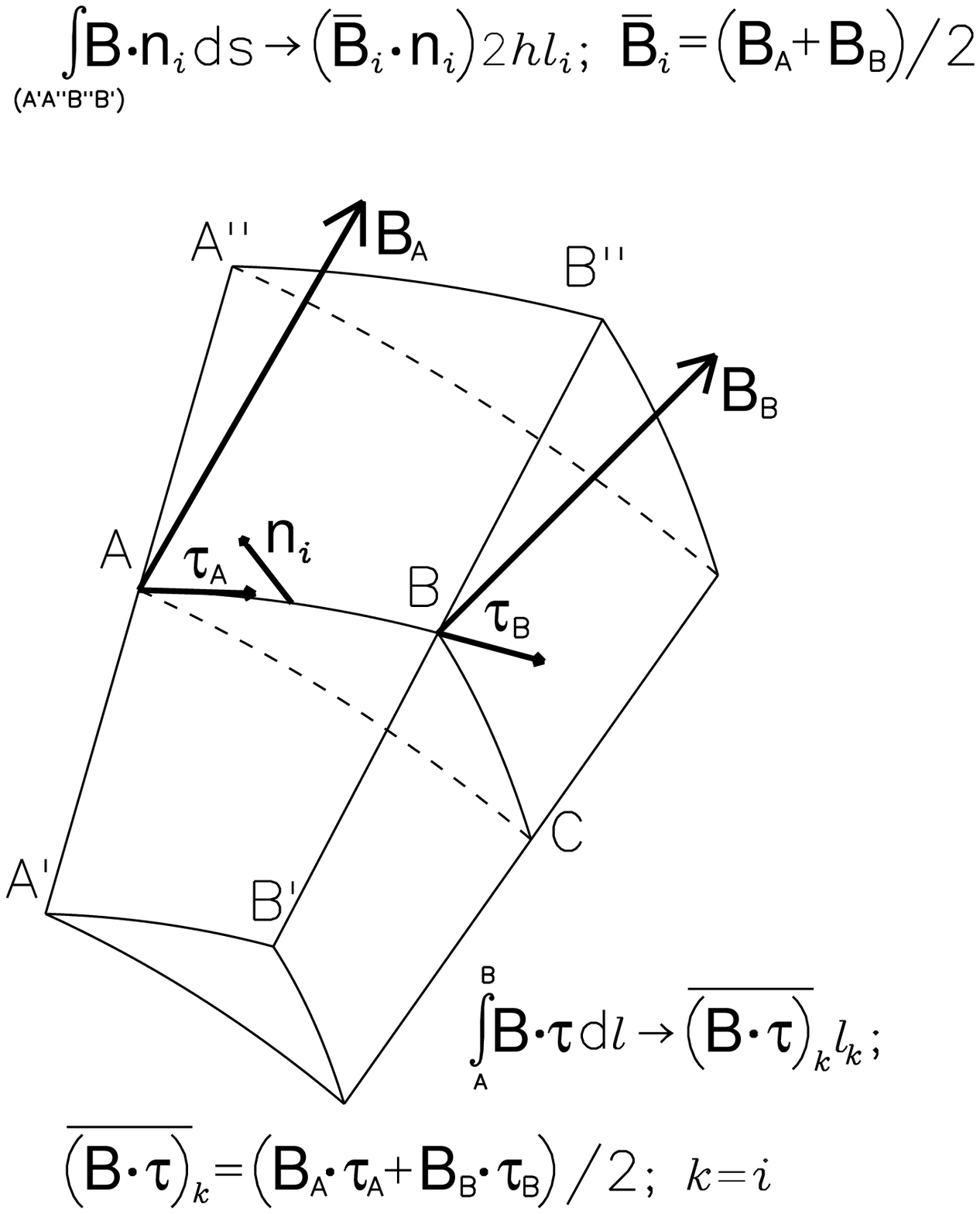}}
\caption{The illustration for Equations (\ref{eq2}) and (\ref{eq3}).
Discrete approximations of integrals through a lateral face of the
spherical prism and through an edge of the spherical triangle with
vertices ($A, B, C$) on the photosphere.} \label{fig1a}
\end{figure}

To construct the local "Energy" function, we apply Gaussian and
Stokes integral theorems to the elementary, arbitrary spherical
triangle whose vertices and edges correspond respectively to three
nearest nodes of the chosen grid and to geodetic intervals jointing
vertices. In this method, desired $\nabla \cdot {\bf B}$ and $J_{n}$
values may be expressed in the form independent of the chosen
coordinate system.

\begin{eqnarray}\label{eq2}
\nabla \cdot {\bf B} \approx \frac{1}{2Sh}\left[
\sum\limits_{i}\left( \overline{{\bf B}}_{i}^{dat}\cdot {\bf
n}_{i}\right) 2hl_{i}+\sum\limits_{j}\left( \overline{{\bf
B}}_{j}^{pot}\cdot {\bf n}_{j}\right) s_{j}\right] \approx \nonumber \\
\approx \frac{1}{S}\sum\limits_{i}\left( \overline{{\bf
B}}_{i}^{dat}-\overline{{\bf B}}_{i}^{pot}\right) \cdot {\bf
n}_{i}l_{i}
\end{eqnarray}

\begin{equation}\label{eq3}
j_{n} \approx \frac{1}{S}\sum\limits_{k}\overline{\left( B\cdot {\bf
\tau }\right) }_{k}l_{k}
\end{equation}
Here $S$ is the spherical triangle area; $h$ is the spherical prism
half-height intersected in the middle by the sphere surface; $l$ is
length geodesic connecting the vertices of the spherical triangle;
index $i$ corresponds to the chosen numbering of lateral faces; $j$
denotes upper and lower faces; ${\bf n}_{i}$ and ${\bf n}_{j}$ are
unit vectors of outer normals of lateral and upper-lower faces
respectively; $s_{j}$ are spherical triangle areas of upper-lower
faces; dashes over field vectors in Equation (\ref{eq2}) correspond
to the mean vector of field from two vertices in the plane
corresponding to the lateral surface; index $k$ in Equation
(\ref{eq3}) numbers consecutively spherical triangle edges; the dash
over the bracket in Equation (\ref{eq3}) denotes the average of two
scalar product fields at vertices of the edge by boundary unit
vectors $\tau $ tangent to the geodetic line, jointing corresponding
pair of vertices directionally ordered in accordance with the
general direction of rotation by the corkscrew rule. To better
understand Equations (\ref{eq2}) and (\ref{eq3}), we give (Figure
\ref{fig1a}) a discrete representation of integrals through a
lateral face of the spherical prism and through an edge of the
spherical triangle in vertices of which the full vector of the
magnetic field under measure is specified.

To construct the divergence of ${\bf B}$, we replace the unknown
field on the upper and lower faces of the spherical prism by the
potential field and then use the strict equality to zero divergence
of the potential field (a similar technique was applied in
\inlinecite {cite02} to construct a divergence difference scheme).

 As a result we use
only nodal field values on the boundary surface to determine local
"energy" $\left( E_{loc}=\left\vert \nabla \cdot {\bf B}\right\vert
+\left\vert j_{z}\right\vert \right) $ Elementary micro-ensemble
"energy" corresponding to a chosen node is defined by the sum of
local "energies" of all triangles for which the chosen node is one
of their vertices.

We propose two release versions of the spherical azimuth ambiguity
removal algorithm: a) on the quasisphere for subsequent
extrapolation and b) on the magnetogram plane to correct only
original magnetograms.

a)  To each node of the quasispherical boundary surface corresponds
a set of radius-vectors in the initial magnetogram coordinate
system. We suppose that in nodes of the uniform quasi-spherical grid
there are vector magnetic data transformed from original, ambiguous
vector magnetograms in the manner of Section \ref{Section2}. We do
not change here the coordinate representation of the vector field
connected with the magnetogram coordinate system. In the
quasi-spherical coordinate system, we calculate a potential vector
field (see Section \ref{Section3}) and express it in terms of
coordinates of the initial coordinate system. Then we start the
spherical azimuth ambiguity removal algorithm with input parameters
(the field of radius-vectors, magnetic field of transformed data,
and calculated potential field).

b) In the magnetogram coordinate system, we calculate radius-vectors
corresponding to nodes of the uniform magnetogram grid. On the
quasi-spherical grid at appropriate spatial resolution, we compute a
potential field and interpolate it back into the magnetogram grid
nodes with simultaneous transformation of its components into the
magnetogram coordinate system. The obtained input data are used for
the disambiguation. This version can be applied to direct
disambiguation of magnetograms without changing positions of nodes
on the magnetogram grid and thus without changing its original
spatial resolution

In conclusion we would like to express our gratitude to the authors
of the original version of the program "Minimum energy" for the free
access to it on the web-site www.cora.nwra.com/AMBIG/. This
alleviated the engineering problem of developing the program code
for the modified version.

\section{Testing with the analytical model of nonlinear force-free field}
      \label{Section5}

\begin{figure}
\centerline{\includegraphics[width=1.\textwidth,clip=]{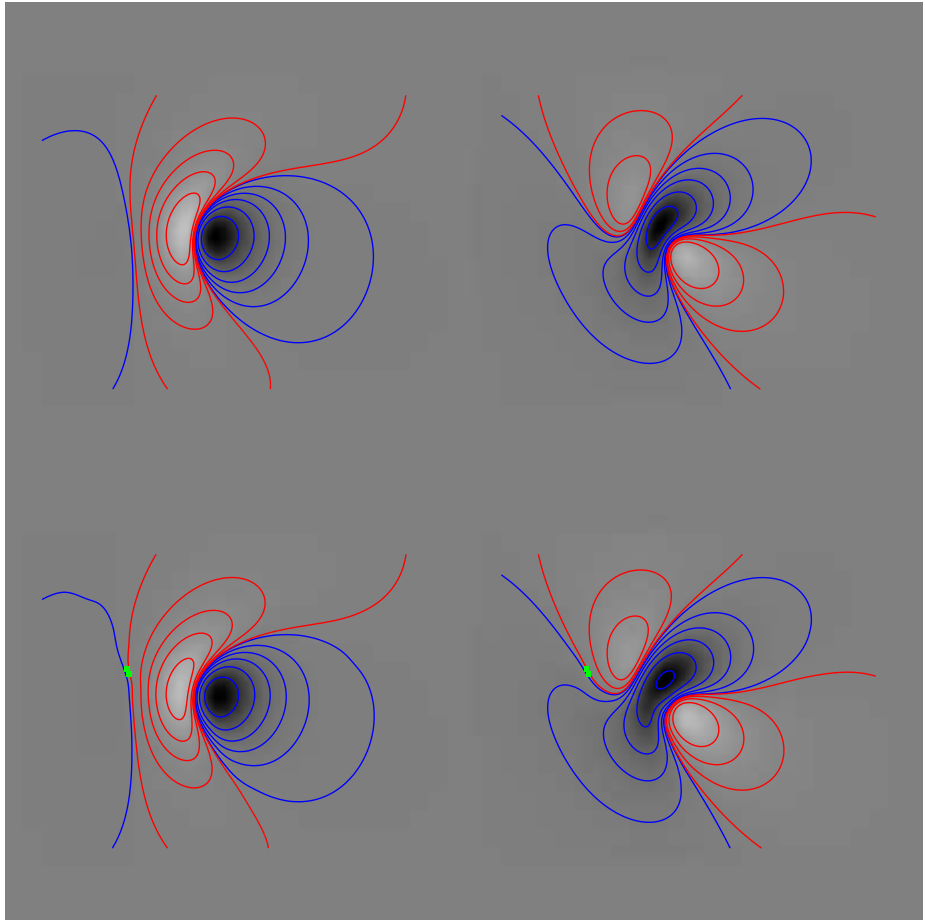}}
\caption{Transverse components of the magnetic field on the
quasi-spherical boundary surface near the limb. On the top panel are
field components of the nonlinear force-free dipole model (Low and
Lou, 1990); on the lower panel are field components prepared for the
extrapolation after the azimuth ambiguity had been resolved by the
spherical minimum energy method. $B_{y}$ (left), $B_{z}$ (right).
The contour lines correspond to a set of field magnitudes, red and
blue lines are positive and negative values respectively. Pixels in
which the angle between vectors of the transverse field in these
magnetograms is over $90^{0}$ are marked by green color. }
\label{fig1}
\end{figure}

\begin{figure}
\centerline{\includegraphics[width=1.\textwidth,clip=]{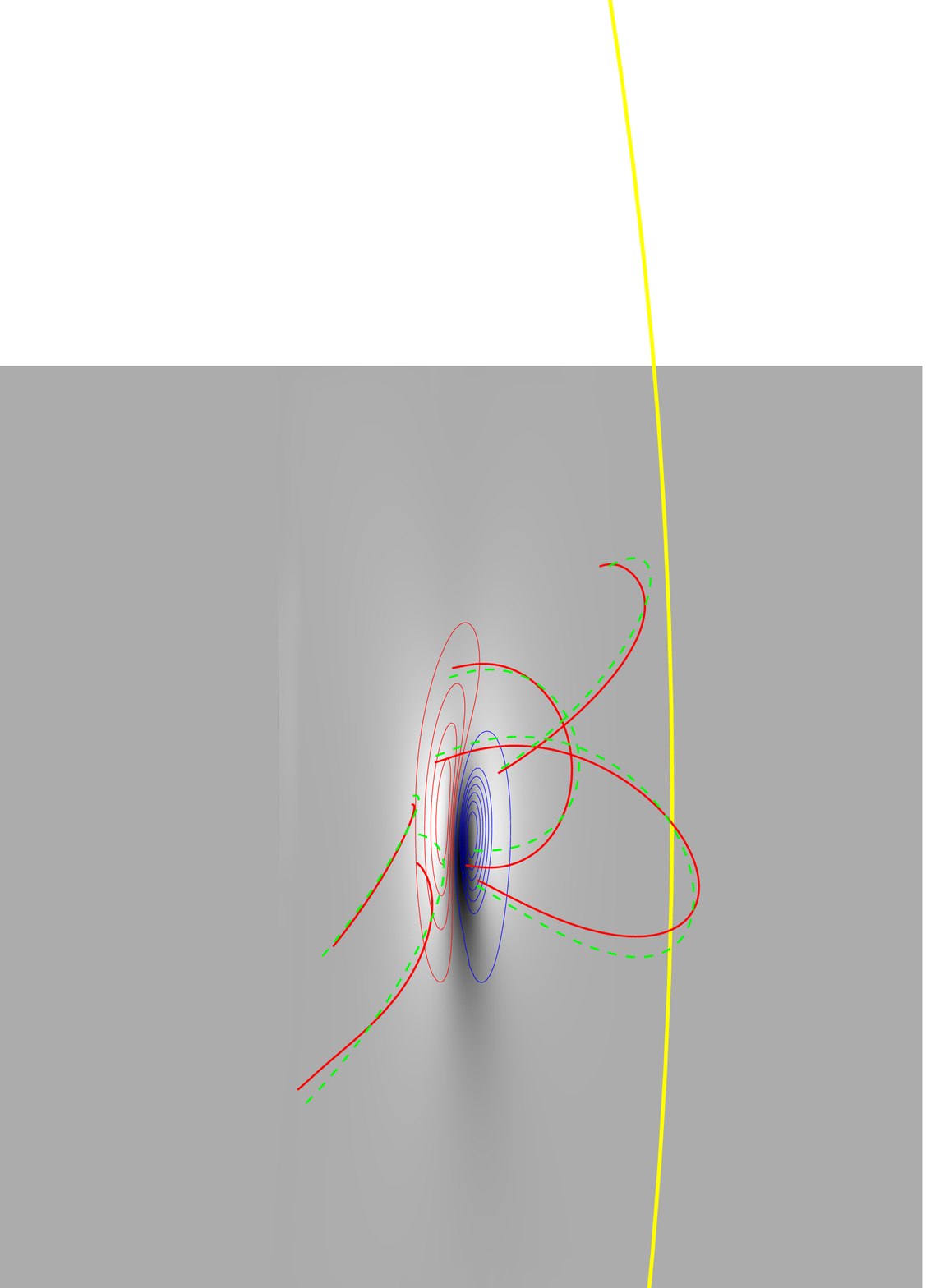}}
\caption{The comparison between calculated field lines (red) with
those (green) of the model field in the plane of the sky of the
solar disk ($[835.8^{\prime \prime}, 1006.2^{\prime \prime}] \times
[-90.0^{\prime \prime}, 80.4^{\prime \prime}]$). The background is
distribution of model field components along the line of sight;
contour lines are distribution of the calculated normal field
component; yellow line is the limb} \label{fig2}
\end{figure}

In the first test, we applied the spherical azimuth ambiguity
removal and extrapolation algorithms to the limb magnetogram
obtained from the analytical model of the nonlinear force-free field
\inlinecite {cite15}. The model's parameters and geometrical layout
were identical to those in \inlinecite {cite01}. The model
magnetogram was calculated for the spherical surface near-limb part.
At the entry, only the field component along the line of sight was
considered unambiguous; the direction of the field transverse to the
line of sight was considered ambiguous ($180^{\circ }$). The test
result is presented in Figures \ref{fig1} and \ref{fig2}. In Figure
\ref{fig1}, the upper pair of transverse components in the given
model field is compared with the lower pair of transverse components
after the azimuth disambiguation. Here we show the transverse field
distribution on the quasi-spherical boundary surface where the
azimuth ambiguity has been resolved. Figure \ref{fig1} presents
 the correct removal of azimuth ambiguity throughout the magnetogram
area. Only in two pixels, the direction of the transverse field of
the disambiguated magnetogram forms an obtuse angle with the model's
field direction. Here the position of the contours shows that the
pixels are in the region of the transverse field's zero module where
the algorithm can not be expected to determine an azimuth correctly.
Figure \ref{fig2} demonstrates effectiveness of the FOM algorithm
for extrapolating the near-limb nonlinear force-free field. This
follows from the correlation between green field lines of the model
field and red ones of the calculated field.

\section{Testings with real magnetograms }
      \label{Section6}
\subsection{Exercises of near center }
      \label{Section6.1}

\begin{figure}
\centerline{\includegraphics[width=1.\textwidth,clip=]{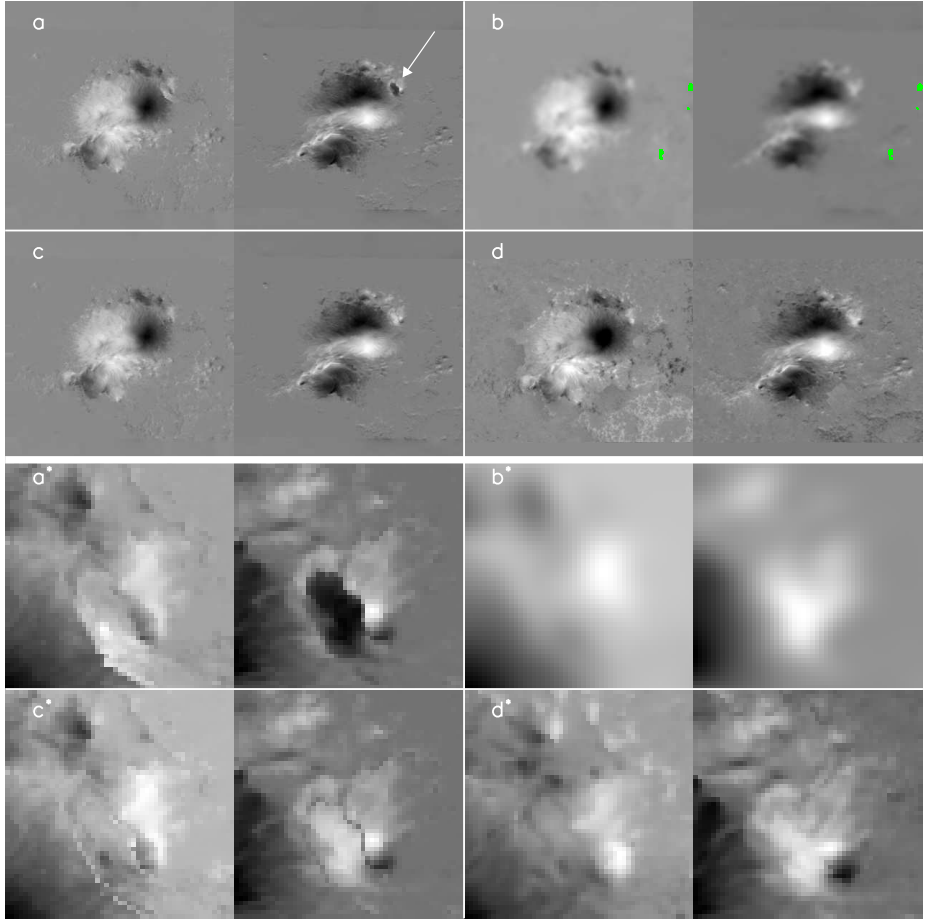}}
\caption{ The transverse magnetograms of $B_{y}, B_{z}$:
Hinode\_SOT\_SP\_20061212\_2030.fits (12 December 2006 20:30 UT)
taken from http://www.lmsal.com/~schryver/NLFFF/ ($a$); the same
magnetograms after the spherical disambiguation on the
quasi-spherical plane with smoothing ($b$); the same magnetograms
after the disambiguation (without smoothing) at high resolution on
the grid uniform in the plane of the sky ($c$); original SOT SP
(Level2) magnetograms (12 December 2006 20:30 UT) after the
disambiguation (without smoothing) at high resolution on the grid
uniform in the plane of the sky ($d$); $a^{\ast }$, $b^{\ast }$,
$c^{\ast }$, and $d^{\ast }$  are parts of corresponding
magnetograms for the fragment of the plane-of-the-sky region
indicated by an arrow. The green regions correspond to noncoincident
directions of the transverse field $> 100$ Gauss on the
quasi-spherical plane of the magnetogram c averaged on the
quasi-spherical plane.} \label{fig3}
\end{figure}

To test the spherical minimum energy method, we compare our
disambiguated magnetograms with the ones (NOAA AR 10930) processed
using the original version of minimum energy method in \inlinecite
{cite14}
\newline
(http://www.lmsal.com/~schryver/NLFFF/Hinode\_SOT\_SP\_20061212\_2030.fits)
\newline
(Figure \ref{fig3}, \emph{a}). The result of the processing of these
magnetograms by the spherical minimum energy method is shown in Fig.
4 \emph{b} and \emph{c}: \emph{b} - the disambiguation made on the
quasi-spherical plane with smoothing over $7\times 7$ nodes (version
(\emph{a}) in Section \ref{Section4}), resultant images are given in
the plane of the sky of the magnetogram; \emph{c} - the
disambiguation made in the original nodes of the plane of the sky of
the initial magnetogram (version (\emph{b}) in Section
\ref{Section4}). One can see that the structure elements of the
transverse field \emph{b} and \emph{ñ} correspond, in the large, to
the original magnetogram a except for the fragment denoted by an
arrow. The marked fragment on the magnetogram a has a region with
sharp boundaries that, in our opinion, indicates misidentification
of the azimuth. In our version (\emph{b}, \emph{ñ}), this part of
the magnetogram looks more natural. Note that under close
examination of the enlarged fragment of our magnetogram ($c^{\ast
}$) at high spatial resolution one can see certain defects in the
form of tortuous contrast channels. In fact, these defects are
artificial and unrelated to algorithm errors in determining the
azimuth. This is supported by the result of the processing we made
at high resolution of the primary magnetogram SOT SP (Level 2) for
12 December 2006 20:30 UT. We have considered two variants of its
preprocessing. In the first case, to comply with the spatial
resolution of the magnetogram \emph{a}, we carried out smoothing
over $3\times 3$ nodes and interpolation onto nodes of the
magnetogram \emph{a} using the technique of artificial Stokes
parameters (in the manner of subsection \ref{Section2.1}).  In the
second case, we conducted direct smoothing and interpolation of the
ambiguous transverse field components. The result of the first
processing is shown in images $d$ ($d^{\ast }$) of Figure
\ref{fig3}. The defects discussed are entirely absent here. In the
second case (no images are given), the contrast channels manifested
themselves in the same form. The data from the magnetogram a must
have been processed in the same manner. It is clear that any
transformations of magnetograms in the ambiguous form break physical
connections between original measurements and thus inevitably affect
disambiguation results.

In closing this subsection let us point out the excellent agreement
between results of the two geometric versions of spherical
disambiguation (\emph{b}, \emph{c}) for the region near the disk
center. Green color marks the regions in which the transverse field
directions of over 100 Gauss in module differ in different
magnetograms (these directions were compared on the quasi-spherical
plane, where the high-resolution field (\emph{c}) was presmoothed to
bring it into line with smoothness of the field \emph{b}).

\subsection{Exercises of near the limb }
      \label{Section6.2}
\begin{figure}
\centerline{\includegraphics[width=1.\textwidth,clip=]{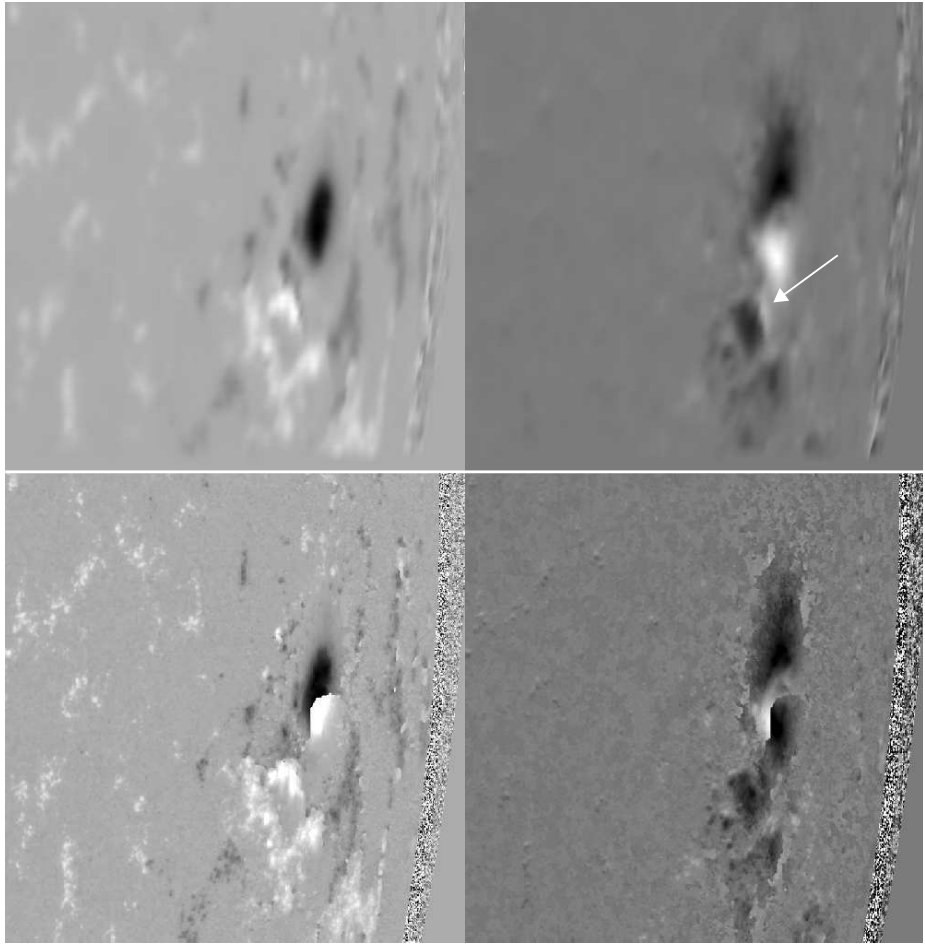}}
\caption{Transverse magnetograms of $B_{y}$, $B_{z}$: SOT SP
(Level2) magnetograms (17 December 2006 0:25:28 UT) after the
spherical disambiguation on the quasi-spherical plane with smoothing
(top); the same magnetograms after the spherical disambiguation
(without smoothing)
 at high resolution on the grid uniform in the plane of the sky
(bottom). The arrow indicates the problem fragment of the top
magnetograms} \label{fig4}
\end{figure}

\begin{figure}
\centerline{\includegraphics[width=1.\textwidth,clip=]{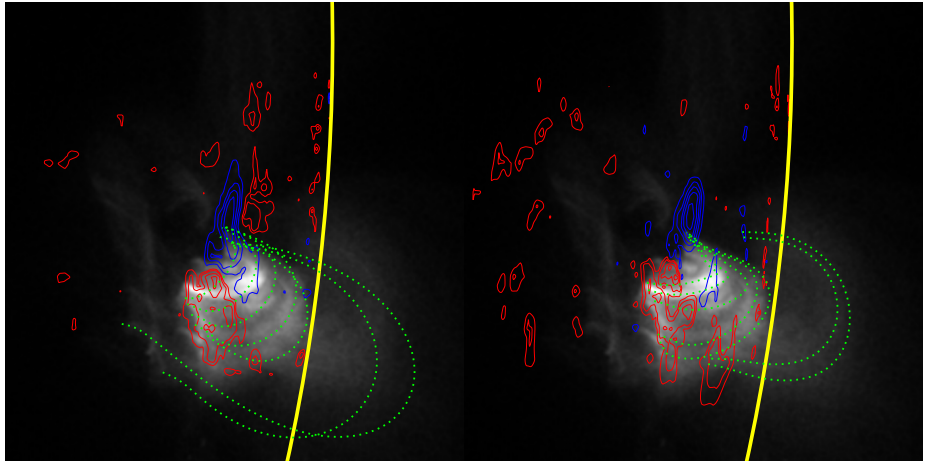}}
\caption{Field lines of the potential (left) and nonlinear
force-free (right) fields. At the bottom is the X-ray image from
{\it Hinode} XRT (17 December 2006 0:24:20.5 UT) with changed
contrast. Contour lines are distribution of the calculated normal
field component; yellow line is the limb } \label{fig5}
\end{figure}

\begin{figure}
\centerline{\includegraphics[height=.5\textheight,clip=]{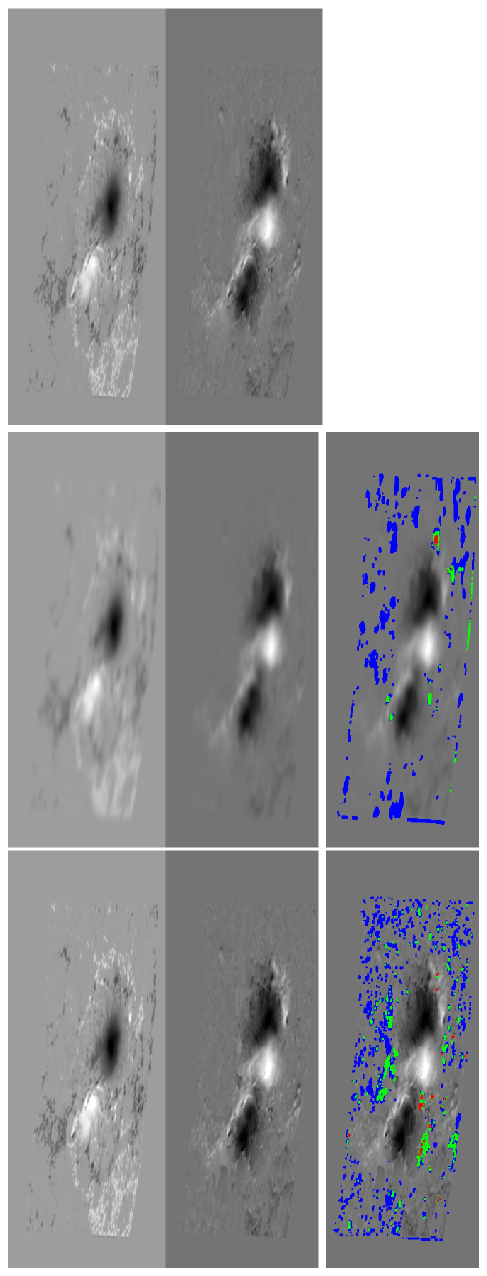}}
\caption{The transverse magnetograms of the model field for 17
December 2006 0:25:28 UT. It was obtained from the magnetogram
\emph{d} of Figure \ref{fig3}: initial model magnetograms (top row);
the magnetograms after the spherical disambiguation on the
quasi-spherical plane with smoothing (middle row); the magnetograms
after the spherical disambiguation (without smoothing) at high
resolution on the grid uniform in the plane of the sky (bottom row).
To the right of the magnetograms on $B_{z}$ component are regions of
noncoincident directions of the transverse field of the
disambiguated and model field magnetograms: blue - $B_{t} > 1$
Gauss, green - $B_{t} > 100$ Gauss, red - $B_{t} > 300$ Gauss;
smoothed and unsmoothed magnetograms are compared respectively with
smoothed and unsmoothed magnetograms of the model field. }
\label{fig6}
\end{figure}

To show the possibilities of correct determination of a transverse
field direction and appropriate extrapolation of near-limb magnetic
regions in the force-free approximation, we have chosen the vector
magnetogram (SOT/SP level 2 "20061217\_002528.fits") of the same
magnetic region NOAA AR 10930 obtained at Space Observatory {\it
Hinode}. The result of the complete cycle of its processing
(disambiguation and extrapolation) is presented by the upper pair of
the smoothed disambiguated magnetograms of the transverse field in
Figure \ref{fig4} and by the system of magnetic field lines of the
calculated field which are superimposed on the image of loop
structures in the X-ray image from Hinode XRT for 17 December 2006
0:24:20.5 UT (Figure \ref{fig5}, right). Due to real loop structures
the force-free simulation model of the magnetic region seems to be
much more preferable than the potential model (Figure \ref{fig5},
left). Figure \ref{fig4} (bottom) shows that the result of the
spherical disambiguation of the initial magnetogram in the plane of
the sky at original spatial resolution (version (\emph{b}) in
Section \ref{Section4}) revealed that the azimuth in the region of
the strong transverse field had been determined incorrectly. The
reason for this might be related to fundamental limitations on work
of the algorithm or to specific character of errors in near-limb
measurements. The first reason may be excluded based on the
secondary analysis we conducted using the artificial model of the
full vector field of the real disambiguated photospheric magnetogram
\footnote{ It is clear that the test made in Section \ref{Section5}
on the smooth analytical model does not adequately reproduce the
complicated character of the real field distribution and gives
insufficient grounds for efficiency of the algorithm}. To construct
this model, we used corrected components of the field of the
magnetogram shown in Figure \ref{fig3} (\emph{d}). The model was
made in the following manner: nodal values of the initial
magnetogram field components were interpolated into nodes of the
uniform grid (with unchanged scale of the initial spatial
resolution) given in the Carrington spherical coordinate system
invariant with respect to rotations; these field components were
transformed into coordinate representation of the Carrington system.
The invariant model simplified production of artificial magnetograms
corresponding to a magnetic region position on the solar disk at a
given time. Figure \ref{fig6} demonstrates quality of the
disambiguation of the artificial magnetogram corresponding to the
exact position of the above real magnetogram. This figure shows
close agreement between the two versions of spherical disambiguation
and the initial artificial magnetogram and thus between each other.
Defects in azimuth correction (color fractions) are in the region of
the insignificant field with relatively low amplitudes; i.e. quality
of the spherical disambiguation can be recognized as adequate.
Comparing these two versions of spherical disambiguation (in area of
defects) shows a better quality of the version with smoothing. This
probably reflects the efficiency of magnetogram noise suppression.

\begin{figure}
\centerline{\includegraphics[width=1.\textwidth,clip=]{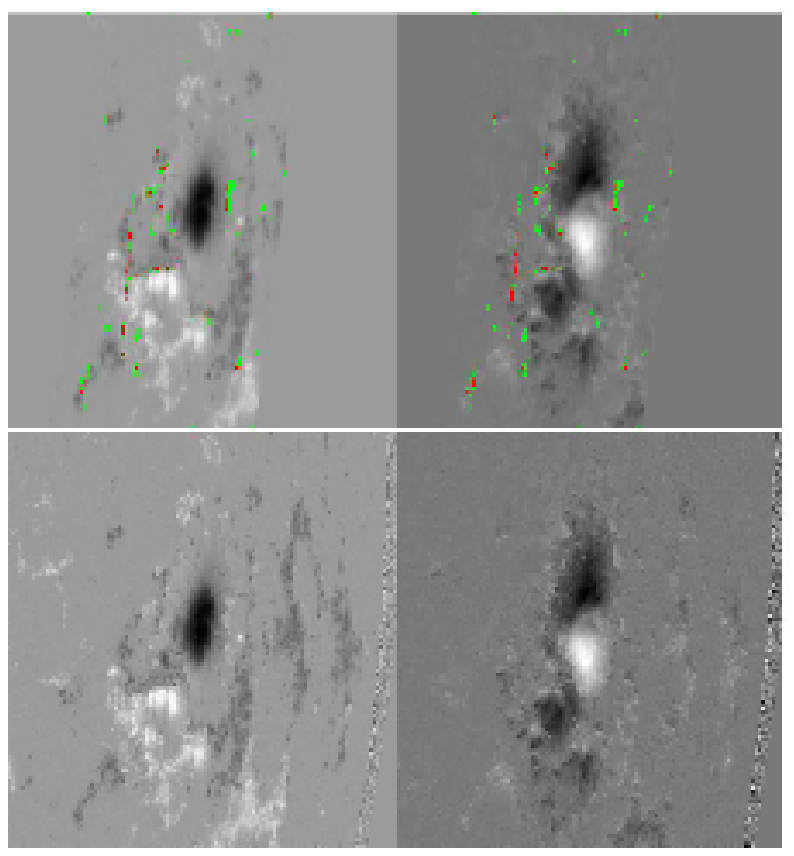}}
\caption{Transverse magnetograms of $B_{y}$, $B_{z}$: SOT SP
(Level2) magnetograms (16 December 2006 12:31:04 UT) after the
spherical disambiguation on the quasi-spherical plane with smoothing
(top panel); the same magnetograms after the spherical
disambiguation (without smoothing)  at high resolution on the grid
uniform in the plane of the sky (bottom panel). The regions of
noncoincident directions of the transverse field on the
quasi-spherical plane of the top magnetograms and the bottom
magnetogram averaged on the quasi-spherical plane are marked by
colors: green - $B_{t} > 100$ Gauss, red - $B_{t} > 300$ Gauss.}
\label{fig7}
\end{figure}

\begin{figure}
\centerline{\includegraphics[width=1.\textwidth,clip=]{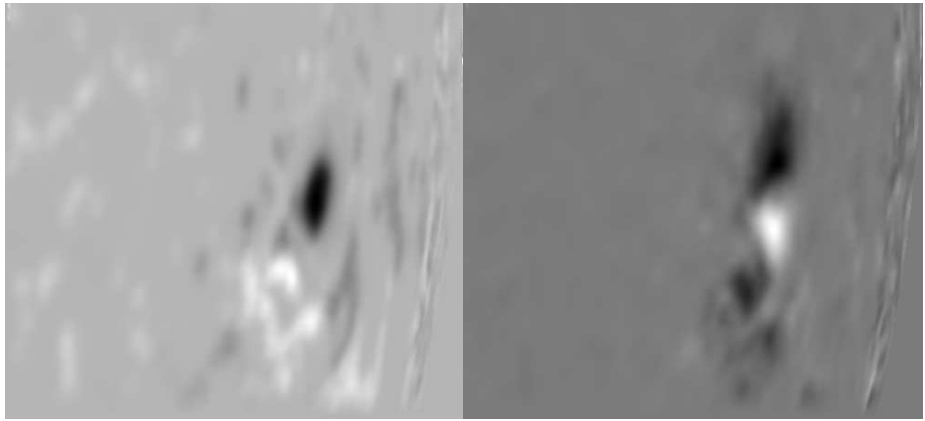}}
\caption{Magnetograms of the upper pair in Figure \ref{fig4} (17
December 2006 0:25:28 UT) after hand correction of the azimuth
direction. } \label{fig8}
\end{figure}

\begin{figure}
\centerline{\includegraphics[width=1.\textwidth,clip=]{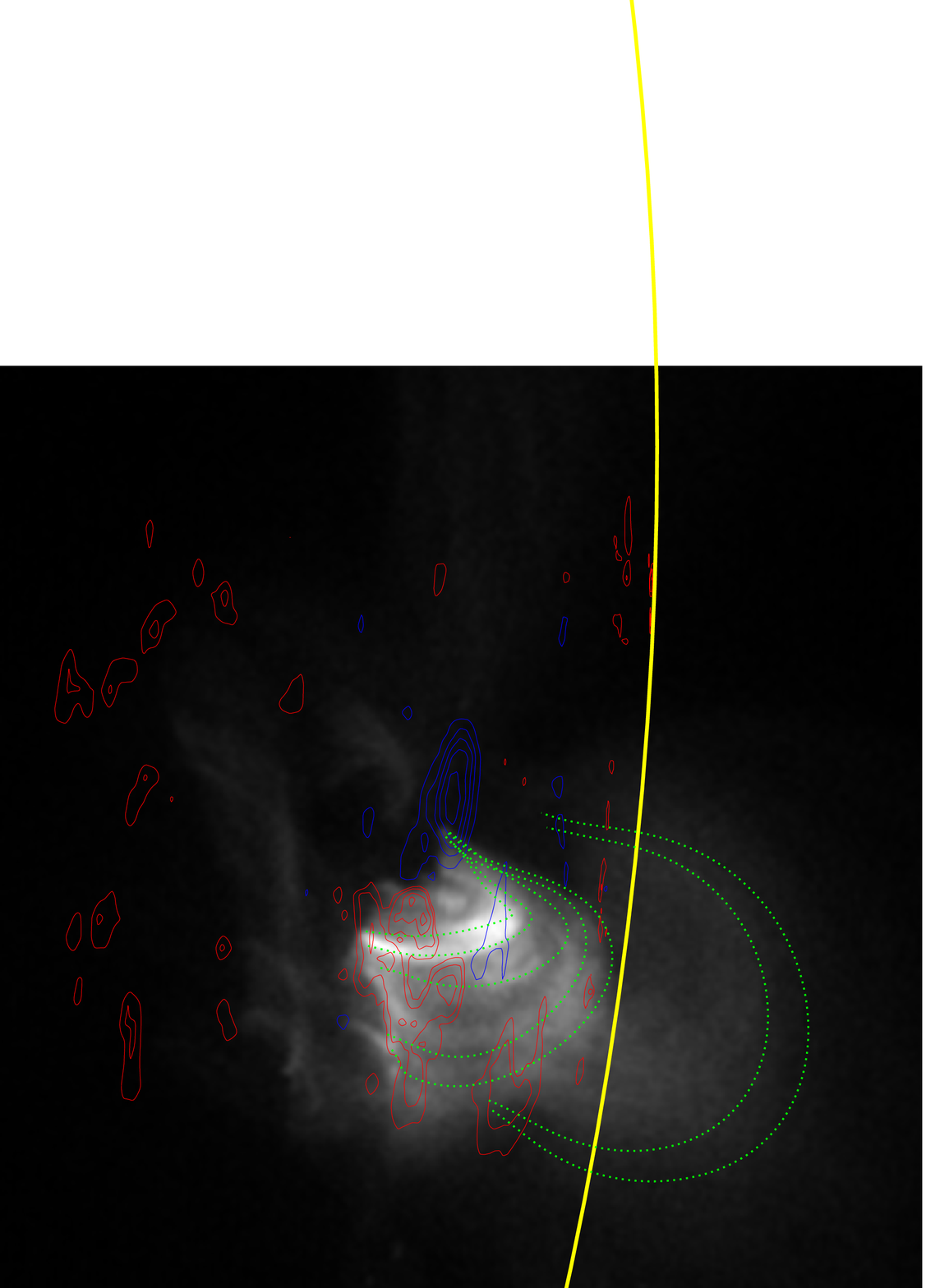}}
\caption{Field lines of the nonlinear force-free field for the
boundary field corresponding to the hand-corrected magnetogram in
Figure \ref{fig8}. At the bottom is the X-ray image from Hinode XRT
(17 December 2006 0:24:20.5 UT) with changed contrast. Contour lines
are distribution of the calculated normal field component; the
yellow line is the limb.} \label{fig9}
\end{figure}

The last discussion raises a question of how far from the limb the
correspondence between these two versions of spherical
disambiguation remains valid. We can get a positive answer to this
question by considering the nearest of the available magnetograms of
this region measured 12 hours before the magnetogram under
discussion (Figure \ref{fig7}). In addition, the obtained transverse
field configuration enables us to detect a small defect in
correcting the azimuth of the top magnetogram in Figure \ref{fig4}
in the fragment indicated by an arrow. This is a small light region
adjacent to the lower negative spot of the $B_{z}$ magnetogram. On
the magnetogram measured earlier (Figure \ref{fig7}), this fragment
is dark and corresponds to the configuration at the time of passage
of the active region near the solar disk center (Fig. 6). This
fragment retaining its polarity so long can not have changed it for
the succeeding 12 hours. This leads us to conclude that from some
distance the limb effect tells on the quality of disambiguation
(though to a lesser extent) not only in the unsmoothed version of
spherical disambiguation. To check the expert assumption about real
direction of the field in the said fragment, we conducted one more
test: made hand correction of the field direction in this fragment
(Figure \ref{fig8}) and extrapolation (Figure \ref{fig9}). A
comparison between Figures \ref{fig5} and \ref{fig9} shows that the
correspondence with the observable loops has not at least gone down.

\section{Conclusions}
      \label{Section7}
We have presented a new algorithm for removing the $180^{\circ }$
azimuth ambiguity. This algorithm is not restricted by the position
of magnetograms on the solar disk. Its application to a real
near-limb region enabled us to test the FOM method of NLFF limb
extrapolation put forward in \inlinecite {cite01}. The findings of
the comparison between the calculated field lines and the real loop
structure confirm the effectiveness of the spherical method for
removing the azimuth ambiguity and the optimization method FOM for
extrapolating nonlinear force-free field the near-limb regions.

\begin{acks}
Hinode is a Japanese mission developed and launched by ISAS/JAXA,
with NAOJ as a domestic partner and NASA and STFC (UK) as
international partners. It is operated by these agencies in
cooperation with ESA and NSC (Norway).
 This work was supported by Grants
Nos. 08-02-92204 and 09-02-00226 of the Russian Foundation for
Fundamental Research; and Lavrentiev Grant of SB RAS 2010-2011.

\end{acks}

\end{article}

\end{document}